\input harvmac
\input amssym.tex

\def\wt{\widetilde{\lambda}}
\def\P{\Bbb{P}}

\lref\BernCU{
Z.~Bern and D.~A.~Kosower,
``Efficient Calculation Of One Loop QCD Amplitudes,''
Phys.\ Rev.\ Lett.\  {\bf 66}, 1669 (1991).
}

\lref\klt{H.~Kawai, D.~C.~Lewellen and S.~H.~H.~Tye, ``A Relation
Between Tree Amplitudes Of Closed And Open Strings,''Nucl.\ Phys.\ B
{\bf 269}, 1 (1986).
}

\lref\WittenNN{
E.~Witten,
``Perturbative gauge theory as a string theory in twistor space,''
arXiv:hep-th/0312171.
}

\lref\RSV{R.~Roiban, M.~Spradlin and A.~Volovich,``A googly amplitude
from the B-model in twistor space,''arXiv:hep-th/0402016.
}

\lref\BerkovitsHG{N.~Berkovits,
``An alternative string theory in twistor space for N = 4
super-Yang-Mills,''arXiv:hep-th/0402045.
}

\lref\ParkeGB{
S.~J.~Parke and T.~R.~Taylor,
``An Amplitude For $N$ Gluon Scattering,''
Phys.\ Rev.\ Lett.\  {\bf 56}, 2459 (1986).
}

\lref\ManganoXK{M.~L.~Mangano, S.~J.~Parke and Z.~Xu,
``Duality And Multi - Gluon Scattering,''Nucl.\ Phys.\ B {\bf 298}, 653
(1988).
}

\lref\mp{M.~L.~Mangano and S.~J.~Parke,
``Multiparton Amplitudes In Gauge 
Theories,''Phys.\ Rept.\  {\bf 200}, 301 (1991).
}

\lref\BerendsME{
F.~A.~Berends and W.~T.~Giele,
``Recursive Calculations For Processes With $N$ Gluons,''
Nucl.\ Phys.\ B {\bf 306}, 759 (1988).
}

\lref\nair{
V.~P.~Nair,
``A Current Algebra For Some Gauge Theory Amplitudes,''
Phys.\ Lett.\ B {\bf 214}, 215 (1988).}

\lref\Penrose{R.~Penrose,``Twistor Algebra,''J.\ Math.\
 Phys.\  {\bf 8}, 345 (1967), 
R.~Penrose,
"Twistor Quantization And Curved Spacetime,"
Int. J. Theor.Phys. {\bf 1} (1968) 61.
}

\Title{\vbox{\baselineskip12pt
	\hbox{hep-th/0402121}
         }}{All Googly Amplitudes from the B-model in Twistor Space}

\centerline{
Radu Roiban${}^\dagger$ and
Anastasia Volovich${}^\ddagger$}

\bigskip
\bigskip

\centerline{${}^\dagger$Department of Physics, University of California}
\centerline{Santa Barbara, CA 93106 USA}

\smallskip

\centerline{${}^\ddagger$Kavli Institute for Theoretical Physics}
\centerline{Santa Barbara, CA 93106 USA}

\bigskip
\bigskip

\centerline{\bf Abstract}

\bigskip

It has recently been proposed that the D-instanton expansion of the
open topological B-model on $\P^{3|4}$ is equivalent to the
perturbative expansion of ${\cal N}=4$ super Yang-Mills theory in
four dimensions.
In this note we extend the results of hep-th/0402016 
and recover the
 gauge theory results for all $n$-point
googly amplitudes by
computing the integral over the moduli space of curves of 
degree $n-3$ in $\P^{3|4}$ .

\Date{February 2004}

\listtoc
\writetoc

\newsec{Introduction}

The construction of efficient computational methods for field theory
scattering amplitudes has benefited substantially from string theory
input. One example is the Bern-Kosower method \BernCU. 
The basis of this technique
is the low energy limit of string scattering amplitudes combined with the
spinor helicity.  This results in enormous calculational
simplifications for loop amplitudes by taking advantage of the
cancelations among Feynman diagrams manifest in the string amplitudes.
Another example is provided by the Kawai-Lewellen-Tye relations
\klt\  among gravity and gauge theory
scattering amplitudes. At tree level these relations are a consequence
of the factorization of the world sheet theory into chiral 
conformal field theories.

Recently Witten proposed a new technique
for computing scattering amplitudes for
${\cal N}=4$ supersymmetric gauge theory in four
dimensions  \WittenNN. In this
construction the gauge theory scattering amplitudes are given by the
D-instanton expansion\foot{
The expressions for the tree level gauge theory amplitudes resulting
from this construction were recently reinterpreted in \BerkovitsHG\ in
terms of a more standard string theory where the D-instanton 
expansion is replaced by the perturbative
expansion.} 
of a topological open string theory  (B-model) whose
target space is the twistor space of the physical space.
The intuition at the base of this conjecture stems from two
sources. The first one is the work of Nair \nair, who showed that the
maximally helicity violating amplitudes (MHV) can be expressed in
terms of correlation functions of 2-dimensional fermionic currents.
The second one is the fact that in a large number of
examples described in \WittenNN\
the scattering amplitudes are supported 
on special curves when transformed
to the twistor space \Penrose\ of Minkowski space.

The  proposal was tested in \WittenNN\ for the case of the MHV
amplitude and in \RSV\ for the 5-point $\overline{\rm MHV}
$\foot{$\overline{\rm MHV}$ 
amplitudes were called ``googly'' in \WittenNN.}
amplitude.
While these results may seem trivial from the gauge theory perspective,
they are certainly not so from the string theory standpoint. Indeed,
the gauge theory relation between MHV and $\overline{\rm MHV}$
amplitudes is just complex conjugation. As we will briefly review in
the next section, the string theory construction does not imply any
simple relation between them, since they are given by integrals over
moduli spaces of curves of different degrees (namely degree one
for MHV and arbitrarily high degree for the conjugate).
Moreover the string
theory construction suggests that the amplitudes may receive
nonvanishing contributions from disconnected curves connected by the
twistor space propagator of the gauge theory fields. Perhaps the most
intriguing part of the result of \RSV\ is that the 5-point
googly amplitude, though supported on degree two curves, 
can be recovered without
contributions from two disconnected degree one curves.

In this short note we will show that the proposal \WittenNN\ 
successfully recovers all $\overline{\rm MHV}$
$n$-point amplitudes. As in the
case of the 5-point amplitude, we find the rather surprising result
that the connected instantons yield the full amplitude without
additional contributions from the disconnected curves.

\newsec{Gluon Amplitudes  and Topological B-model}

The gauge theory results for the $\overline{\rm MHV}$   (googly)
amplitudes 
(amplitudes with $n-2$ negative helicities and $2$ positive
helicities) is well known: 
they are complex conjugate to MHV amplitudes
(amplitudes with $n-2$ positive helicities and $2$ negative
helicities)\ParkeGB.
In the spinor helicity
notation $p_{a{\dot a}}=\lambda_a{\tilde\lambda}_{\dot a}$ they 
are given by the simple expression
\eqn\googly{\eqalign{
&{A}_{\rm googly}
(\lambda,\wt,\eta)
= i g^{n-2} (2\pi)^4 \delta^4
\left( \sum_{i=1}^n \wt_i^{\dot{a}} \lambda_i^a \right)
\cr
&\qquad\qquad\qquad\qquad\qquad\times
\int d^{4 n} \psi \, \exp\left[{i \sum_{i=1}^n \eta_{iA}
\psi_i^A}\right]
\delta^8 \left( \sum_{i=1}^n \wt_i^a \psi_i^A
\right)
\prod_{i=1}^n {1 \over [i,i+1]},
}}
where the spinor product $[i,j]$ defined as $[i,j] \equiv 
[\wt_i,\wt_j] = \epsilon_{\dot{a} \dot{b}}
\wt_i^{\dot{a}} \wt_i^{\dot{b}}.$

Recently Witten \WittenNN\ conjectured that tree 
level $n$-particle gauge theory amplitudes
are supported on curves of degree $d=q-1,$ where
$q$ is the number of negative helicity gluons.
He  also proposed  that gluon scattering amplitudes
in  ${\cal N}=4$ SYM
can be computed from the 
open topological B-model on the supertwistor space 
$\P^{3|4}$  in the presence of D5 and D1 branes. 
In this construction SYM fields are realized as the
excitations of the open strings on a D5 brane and the scattering
amplitudes are computed by ``integrating out'' the D1 brane fields.
The introduction of  D1 branes breaks most of the isometries of
$\P^{3|4}$ and in order to restore them one has to
integrate over all possible configurations of
the D1 branes whose genus and degree are determined by the
amplitude of interest.
 
Putting together all the details and parametrizing the 
moduli space of genus zero and degree $d$ curves in terms of degree
$d$ maps from $\P^1$ into $\P^{3|4}$, we find that the 
master formula for the tree-level contribution to $n$-gluon scattering
from instantons of degree $d$ (relevant when there are $d+1$ negative  
helicity gluons) is \WittenNN, \RSV\
\eqn\main{
B(\lambda,\mu,\psi)
=\int { d^{4d+ 4} a\ d^{4d+4} \beta \ d^n \sigma
\over {\rm vol}(GL(2))}
\prod_{i=1}^n
{ 1 \over
\sigma_i - \sigma_{i+1}}
\delta^3 \left( {z_i^I \over z_i^J }
- { P^I(\sigma_i)
\over
P^J(\sigma_i)}\right)
\delta^4\left( {\psi_i^A \over z_i^J}
-  { G^A(\sigma_i)
\over
P^J(\sigma_i)}
\right),
}
where 
\eqn\aaa{
z^I=P^I(\sigma) = \sum_{k=0}^d a_k^I \sigma^k~~, \qquad
\psi^A=G^A(\sigma) = \sum_{k=0}^d \beta_k^A \sigma^k~~,
}
and $z^I = (z^0,z^1,z^2,z^3) =
(\lambda^1,\lambda^2,\mu^1,\mu^2)$ are the homogeneous bosonic
coordinates on $\P^{3|4}$, $\psi^A$, $A=1,2,3,4$
are the fermionic coordinates, $\sigma$ is the inhomogeneous
coordinate on $\P^1$ and $I\ne J$.  This equation was used in \WittenNN\ 
to recover the MHV amplitudes from $d=1$ curves and in \RSV\ to
recover the $n=5$ $\overline{\rm MHV}$ amplitudes from $d=2$. 
In the next section we recover $n$-point $\overline{\rm MHV}$ 
amplitudes from 
\main.

\newsec{The B-Model Calculation}

In this section we evaluate the Fourier transform $\widetilde{B}$
of the B-model amplitude \main\ with respect to $\mu$ for 
the case $n=d+3$, which is
relevant to the scattering of $n-2$ negative and 2 positive helicity
gluons in YM theory. In the following, we will keep $d$ arbitrary in
the equations which are independent of the relation between the number
of external legs and the degree of the curve.

To simplify the equations, we will analyze separately the bosonic and
the fermionic parts of equation \main. Choosing the index $J=0$ in 
equation
\main\ and Fourier transforming $\mu\rightarrow \wt$ the bosonic 
part of the amplitude, we find
\eqn\aaa{
\eqalign{
\widetilde{B}(\lambda, \wt) =
&\int { d^{4(d+1)} a\,d^n \sigma
\over {\rm vol}(GL(2))}
\,J_0\,
\left[
\prod_{i=1}^n
\delta \left( {\lambda_i^2 }
- { P^1(\sigma_i) \over P^0(\sigma_i) } \right)
\right]
\exp \left[ i \sum_{i=1}^n
\sum_{k=0}^d
{\epsilon_{\dot{a} \dot{b}}
\wt_i^{\dot{a}} a_k^{\dot{b}} \sigma_i^k  \over P^0(\sigma_i) }
\right],
}}
where
\eqn\aaa{
J_0 = \prod_{i=1}^n { 1 \over
\sigma_i - \sigma_{i+1}}~~.
}
As in \RSV\  we absorbed $\lambda^1$ in $\lambda^2$ and $\mu^{\dot a}$.
We will reinstate it at the end of the calculations through the
rescaling  $\lambda^2\rightarrow \lambda^2/\lambda^1$ and 
${\tilde \lambda}^{\dot a}\rightarrow {\tilde \lambda}^{\dot a}
\lambda^1$.

The first step is to fix the GL(2) symmetry by setting
the variables
$a_0^0$, $\sigma_1$, $\sigma_2$ and $\sigma_3$ to some arbitrary
values at the cost of introducing the Jacobian
\eqn\aaa{
J_1 = a_0^0 V_{123}, \qquad V_{123} \equiv (\sigma_1 - \sigma_2)(\sigma_2
- \sigma_3)(\sigma_3 - \sigma_1)~~.
}

The integral over the $2(d+1)$ $a_k^{\dot{a}}$ moduli is trivial and
gives 
\eqn\six{
\widetilde{B}=
\int  d^{d}a\,d^{d+1}b\, d^{n-3}\sigma_i\,
J_0 J_1
\left[
 \prod_{i=1}^n
\delta \left( {\lambda_i^2 }
- { B_i \over A_i } \right)
\right]
\prod_{k=0}^d
\delta^{2} \left(
\sum_{i=1}^n { \wt_i^{\dot{a}}  \sigma_i^k \over
A_i} \right)~~.
}
Here we have parametrized the remaining bosonic moduli
by $a_k$ (with $a_0=a_0^0$ unintegrated) and $b_k$, with
\eqn\aaa{
A_i = \sum_{k=0}^d a_k \sigma_i^k~~, \qquad
B_i = \sum_{k=0}^d b_k \sigma_i^k~~.
}

A simple counting reveals that there are enough
$\delta$ functions to fix all the integration variables to a discrete set
of values. The first goal is to find them. It turns out that, for
$d=n-3$, there is a unique set of $\sigma_i$ and $A_i$ which satisfies
the constraints imposed by the last $2 (d+1)$ delta functions. 
To find them we notice that the corresponding equations
are linear in the ratio $r_j={A_1/ A_j}$ and their solution is
\eqn\solr{
r_j={A_1 \over A_j}={[k,1] \over [k,j]}
{(\sigma_1-\sigma_2)(\sigma_1-\sigma_3) 
\over (\sigma_j-\sigma_2)(\sigma_j-\sigma_3)}
\prod_{p=4, p\neq  k,j}^n {(\sigma_1-\sigma_p) \over
(\sigma_j-\sigma_p)}~~~~~{\rm for}~~j\ne k~~,
}
where $j= 4,\dots, n$ and $k$ can be chosen to be any number from $4$
to $n$. Since the result should not depend on the choice of $k$ we
find that the following equation must be satisfied
\eqn\urav{
{[k_1,1] \over [k_1,j]} {\sigma_1-\sigma_{k_2} \over
\sigma_j-\sigma_{k_2}}= 
 {[k_2,1] \over [k_2,j]} {\sigma_1-\sigma_{k_1} \over
\sigma_j-\sigma_{k_1}}~~~~~(\forall)~~j\ne k_1\ne k_2~~. 
}
Using the same reasoning for ${A_2 / A_j}$ and ${A_3 / A_j}$
we find two more equations for $\sigma_{k_1},$
$\sigma_{k_2}$ and $\sigma_j$. They are similar to \urav\ except that
the index $1$ is replaced by $2$ and $3$, respectively.
These three equations have a unique solution 
\eqn\solsig{
\sigma_j={ \sigma_2 \sigma_3 [j,1][2,3]
+ \sigma_1 \sigma_2 [j,3][1,2]+
 \sigma_1 \sigma_3 [j,2][3,1]
\over
 \sigma_1 [j,1][3,2]+
 \sigma_3 [j,3][2,1]+
 \sigma_2 [j,2][1,3]
}~~.
}

The next step is to expose the momentum conservation constraint and
cast the remaining constraints into a more useful form. To
this end we make use of the remarkable identity
\eqn\remarkable{
\eqalign{
&\left[ \prod_{i=1}^n
\delta \left( {\lambda_i^2 }
- { B_i \over A_i } \right)
\right]
\prod_{k=0}^d
\delta^{2} \left(
\sum_{i=1}^n { \wt_i^{\dot{a}}  \sigma_i^k \over
A_i} \right)
\cr
&
\quad= 
J_2
\delta^4 (p)
\left[
\prod_{i=1, i\neq 2,3}^n
\delta \left( {\lambda_i^2 }
- { B_i \over A_i } \right)
\right]
\prod_{k=1}^d 
\delta (\sum_{i=1}^n  {[i,2] \sigma_{i3} \sigma_i^{k-1} \over A_i})
\delta (\sum_{i=1}^n  {[i,3] \sigma_{i2} \sigma_i^{k-1} \over A_i})~~,
}}
where
\eqn\aaa{
J_2 = A_2 A_3 A_1^{2 d} [2,3]^{d+1}~~~~{\rm and} 
~~~~\sigma_{ij} \equiv(\sigma_i-\sigma_j)~~.
}
As promised, the first delta function enforces the momentum conservation
$\delta^4 (p)=\delta^4(\sum_{i=1}^n \wt_i^{\dot{a}} \lambda_i^{\dot{a}})$
after restoring $\lambda^1_i$ dependence.

The integrals over the $(d+1)$ $b$ moduli can be easily performed,
given the fact that they appear linearly in the equation \remarkable.
The resulting factor is
\eqn\aaa{
J_3 = { \prod_{i=1, i\neq 2,3}^n  A_i \over V_{14 \cdots n}}~~,
}
where $V_{14 \cdots n}$ is the Vandermonde determinant of
$\sigma_1, \sigma_4, \cdots \sigma_n.$

The last step is to perform the integrals
over $a_i$ and $\sigma_i$. Perhaps the easiest way to do this is to 
make use of the fact that the arguments of the last delta functions are
linear in $A_1/A_i$. Changing the integration variables from $a_i$ to 
$r_i=A_1/A_i$ with $i=4,\dots,n$ introduces the Jacobian
\eqn\aaa{
J_4={ A_1^{4-n}\prod_{i=4}^n  A^2_i \over V_{14 \cdots n}}
} 
Using the $d=n-3$ $\delta$ functions in the equation
\remarkable, the integrals over the ratios $r_i$ as well as over the
positions of the fermionic currents $\sigma_i$ yield
\eqn\aaa{
J_5=\left|
\matrix{
[2,i] \sigma_{i3 }\sigma_i^{k-1} &
[2,i] r_i (k\sigma_{i3}+\sigma_3) \sigma_i^{k-2}
\cr
[3,i] \sigma_{i2 }\sigma_i^{k-1} &
[3,i] r_i (k\sigma_{i2}+\sigma_2) \sigma_i^{k-2}
}
\right|^{-1}}~~.

Multiplying  the rows in the first block by
${[1,3] \sigma_{21} \over [2,1] \sigma_{13}}$
and subtracting them from the rows of the second block while using the
equation \solr\ leads, after some algebra, to the last piece of the
bosonic integrals 
\eqn\aaa{
J_5={1 \over \sigma_{23}^d V_{4, \cdots n}^2 \prod_{i=4} ^n([2,i] [3,i]
r_i)}~~ .
}

The contribution of the fermionic moduli can be easily computed with
the result
\eqn\fermion{
F \equiv
\int d^{4(d+1)} \beta \prod_{i=1}^n
\delta^4 \left( \psi_i^A - \sum_{k=0}^d
{ \beta_k^A \sigma_i^k \over A_i}\right)=
\left(
V_{14 \cdots n}
\over
A_1 A_4 \cdots A_n [2,3]
\right)^4
 \delta^8
\left( \sum_{i=1}^n \wt_i^{\dot{a}} \psi_i^A\right).
}

Evaluating the result of the bosonic integrals $J_0 J_1 J_2 J_3 J_4 J_5$
on the solution \solsig\ and multiplying by the result of the
integral over the fermionic moduli \fermion\ while restoring the
$\lambda_i^1$ dependence by rescaling $\lambda_i^2$, $\wt_i^{\dot\alpha}$
and $\psi^A$ yields the full B-model amplitude
\eqn\aaa{
\widetilde{B}(\lambda,
\wt,\psi)= \delta^4\left( \sum_{i=1}^n \wt_i^{\dot{a}} \lambda_i^a
\right) \delta^8 \left( \sum_{i=1}^n \wt_i^{\dot{a}}
\psi_i^A \right) {1 \over \prod_{i=1}^n [i,i+1]}~,
}
in agreement with \googly\ after the necessary fermionic 
Fourier transform.

It is worth pointing out the difference between the gauge theory and
the string theory derivation of the MHV and $\overline{\rm
MHV}$ amplitudes. The string theory proposed in \WittenNN\ directly
reproduces the n-point $\overline{\rm MHV}$ amplitude while 
from the gauge theory standpoint they are obtained by
solving certain recurrence relations \refs{\ParkeGB, \ManganoXK,
\BerendsME, \mp}. It may turn out that, in general,
the most efficient way of proving the proposal \WittenNN\  
is to derive the gauge theory recursion relations from the string
field theory of the topological B-model on the twistor space. It is
however our hope that a generalization of the methods described here
will prove to be a useful tool in the explicit computation of gauge
theory scattering amplitudes.

\bigbreak\bigskip\bigskip
\centerline{\bf Acknowledgments}\nobreak

We have benefited from helpful discussions
with  D. Gross,  D. Kosower, W. Siegel,  E. Witten
and especially M. Spradlin.
This work was supported in part by the National Science Foundation
 under PHY00-98395 (RR) and PHY99-07949 
(AV), as well as by the DOE under Grant
No.~91ER40618 (RR).  Any opinions, findings, and conclusions or
recommendations expressed in this material are those of the authors
and do not necessarily reflect the views of the National Science
Foundation.

\listrefs

\end